\documentclass[proof]{pasj00}
\draft

\begin{document}

\SetRunningHead{K. Enya et al.}
{Manufactureing of coronagraphic binary pupil masks}

\title{
Comparative study of manufacturing techniques
for coronagraphic binary pupil masks:
masks on substrates and free-standing masks
}


\author{Keigo \textsc{Enya},$^1$ Kanae \textsc{Haze},$^1$ 
Takayuki \textsc{Kotani},$^1$ and Lyu \textsc{Abe}$^2$}  

\affil{$^1$Department of Infrared Astrophysics, 
Institute of Space and Astronautical Science,\\ 
  Japan Aerospace Exploration Agency
  Yoshinodai 3-1-1, Sagamihara, Kanagawa 229-8510}
\affil{$^2$
Laboratoire Universitaire d'Astrophysique de Nice, 
UMR 6525, Parc Valrose, F-06108 Nice, France}


\email{enya@ir.isas.jaxa.jp}



\maketitle


\begin{abstract}
We present a comparative study of the manufacture of binary pupil 
masks for coronagraphic observations of exoplanets. 
A checkerboard mask design, a type of binary pupil mask design, 
was adopted, and identical patterns of the same size were used 
for all the masks in order that we could compare the differences 
resulting from the different manufacturing methods. 
The masks on substrates had aluminum checkerboard patterns 
with thicknesses of 0.1/0.2/0.4/0.8/1.6$\mu$m constructed on 
substrates of BK7 glass, silicon, and germanium using 
photolithography and chemical processes.
Free-standing masks made of copper and nickel with thicknesses 
of 2/5/10/20$\mu$m were also realized using photolithography 
and chemical processes, 
which included careful release from the substrate 
used as an intermediate step in the manufacture.
Coronagraphic experiments using a visible laser were carried 
out for all the masks on BK7 glass substrate and the free-standing masks. 
The average contrasts were 8.4$\times10^{-8}$, 1.2$\times10^{-7}$, 
and 1.2$\times10^{-7}$ for the masks on BK7 substrates, 
the free-standing copper masks, and the free-standing nickel masks, 
respectively.
No significant correlation was concluded between the contrast 
and the mask properties.
The high contrast masks have the potential to cover the needs 
of coronagraphs for both ground-based and space-borne telescopes 
over a wide wavelength range. Especially, their application 
to the infrared space telescope, SPICA, is appropriate.

\end{abstract}


\section{Introduction}

The direct detection and spectroscopy of exoplanets is expected 
to play an essential role in the understanding of how planetary 
systems were born, how they evolve, and, ultimately, 
in finding biological signatures on these planets. 
For the direct observation of exoplanets, the enormous contrast 
in luminosity between the central star and the planet is a critical difficulty. 
For example, the contrast between the sun and the earth observed 
from outside is $\sim$10$^{-10}$ at visible light wavelengths 
and $\sim$10$^{-6}$ in the mid-infrared wavelength region, 
respectively(\cite{Traub2002}).
Therefore, the number of exoplanets detected directly is quite 
a lot smaller than the number of those detected by other methods
(e.g., \cite{Mayor1995}; \cite{Charbonneau2000}), 
though the first direct observation was finally achieved
(e.g., \cite{Marois2008}; \cite{Kalas2008}; \cite{Lagrange2010}).
Coronagraphs, which were first developed for solar 
observations (\cite{Lyot1939}), is special optics to reduce the contrast.
It is considered that advanced coronagraphs have the potential 
to make possible further extended direct observations of exoplanets.

Among the various current coronagraphic methods, coronagraphs 
using binary pupil masks have some advantages,
and has been studied(\cite{Jacquinot1964}; \cite{Spergel2001};
\cite{Vanderbei2003a}; \cite{Vanderbei2003b}; 
\cite{Vanderbei2004}; 
\cite{Kasdin2005a}; \cite{Kasdin2005b}; 
\cite{Belikov2007}; \cite{Enya2007}; 
\cite{Enya2008};
\cite{Haze2009}; 
\cite{EnyaAbe2010};
\cite{Carlotti2011}; \cite{Haze2011}; \cite{Enya2011a}; 
\cite{Haze2012}).
The function of a binary pupil mask coronagraph to produce a high 
contrast point spread function(PSF) is so less sensitive to 
wavelength (except the effect of scaling the size of the PSF), 
and also be quite less sensitive to telescope pointing errors 
than other coronagraphs.  
Simplicity is another advantage of this optics.
Because of these advantages, the use of a binary pupil mask 
coronagraph is being considered (e.g., \cite{Enya2011b}) for 
the Space Infrared Telescope for Cosmology and Astrophysics(SPICA) 
mission (e.g., \cite{Nakagawa2009}).

For the development of a binary pupil mask coronagraph, 
both free-standing masks and masks constructed on substrates 
are possible. In laboratory demonstration experiments, 
a high contrast of 6.7$\times 10^{-8}$ was achieved with a high 
precision mask constructed on a glass substrate by electron beam 
lithography(\cite{Enya2008}). 
On the other hand, masks on substrates have undesirable properties. 
The substrates give rise to transmittance losses, and the applicable 
wavelength of the coronagraph is limited by the substrate material. 
Multiple reflections at the front and back surfaces of the mask 
are another disadvantage.

Considering this background, we carried out a comparative study 
of mask manufacturing processes. 
Aluminum(Al) mask patterns of various thicknesses were manufactured 
on substrates of BK7 glass, silicon(Si), and germanium(Ge).
Free-standing masks made of copper (Cu) and nickel (Ni) of 
various thicknesses were also manufactured. 
The design of the mask pattern was common to all the masks 
manufactured so that we were able to carry out a systematic 
comparison of the coronagraphic performance focusing on the 
differences in the manufacturing processes. 
In this work, we set the primary goal contrast to be 10$^{-6}$ 
because of the need to observe exoplanets using space infrared telescopes. 
The design, manufacture, and the results of laboratory tests 
of the coronagraphic performance are presented in the following sections.

\section{Mask Design}

\subsection{Checkerboard Pattern}

Among the various binary pupil masks for coronagraphs, 
we chose the checkerboard mask for the following reasons:
First, the topology of the checkerboard mask design essentially 
guarantees the possibility that it can be made as a free-standing mask. 
Second, the pattern consisting of many rectangular apertures 
formed of orthogonal straight lines is suitable for our manufacturing processes, 
rather than other masks consisting of apertures with 
smooth curves (e.g., Gaussian shaped masks).
The design of the checkerboard pattern adopted in this work 
is shown in Fig.\ref{fig1}, in which four dark regions(DRs), DR1--DR4, 
are produced around the core of the PSF.
The contrast, Inner Working Angle($IWA$), and Outer Working Angle($OWA$) 
in the design are $10^{10}$, 5.4$\lambda/D$, and 50$\lambda/D$, 
respectively, in which $\lambda$ is the wavelength and $D$ is 
the length of the diagonal of the whole checkerboard pattern.
It should be noted that the LOQO optimizer presented 
by \citet{Vanderbei1999} was used for optimizing the design.

\subsection{Variation}

Table.\ref{table1} summarizes the parameters of 
the various masks used in this work.
For the material of the substrate, BK7 glass, Si, 
Ge were adopted.
Masks on BK7 glass substrates are convenient as they can 
be tested with a visible light source at ambient temperature in air. 
Such tests for the masks on Si or Ge substrates are not 
possible because Si and Ge are opaque in the visible light 
wavelength region.
On the other hand, masks on Si and Ge substrates can 
potentially be used in the mid-infrared wavelength region.
The mask pattern was formed in aluminum(Al) of various 
thicknesses on all of these substrates.
The thicknesses were 0.1/0.2/0.4/0.8/1.6$\mu$m.
The free-standing masks were manufactured in various 
thicknesses of either Cu or Ni.
We considered high precision manufacture with Cu is 
more common and established than Ni.
On the other hand, Ni is physically tougher and more 
resistant to oxidation than Cu.
The thicknesses of the free-standing masks were 2/5/10/20$\mu$m.

\subsection{Geometry}

The geometry of the masks is shown in Fig.\ref{fig2}.
The geometry of the substrates was determined by the 
availability of those suitable for our manufacturing process.
As a result, $\phi$ 30mm BK7 glass substrates, and 
50mm-square Si and Ge substrates were selected.
For all the masks on substrates, the checkerboard 
pattern was located at the center of the substrate. 
The whole geometry of the free-standing masks was 
designed in order to realize actually “free-standing”
and to make realistic holding possible. 
As the result, we adopted a 50mm-square design with 
thicker holding area around the checkerboard pattern area.
A goal thickness of the holder area in design is $\geq$100$\mu$m.

\section{Manufacturing processes}

\subsection{Photo-mask for micro-structure patterning}

First, a photo-mask was manufactured on a special glass 
substrate, which was used the master
pattern of micro-structure 
of all the coronagraphic masks manufactured in this work. 
Details of the manufacturing process for the photo-mask 
are shown below, and also in Fig.\ref{fig3}.

\begin{enumerate}

\item A special 4-inch (101.6mm) square glass plate is used as the substrate

\item The substrate is coated with $Cr+Cr_{2}O_{3}$ (0.1 $\mu$m thickness) 
by sputtering, a process in which a thin layer of metal is deposited 
by momentum exchange between energetic ions in 
a plasma and the atoms in a target material.

\item A 0.5$\mu$m thick layer of photoresist is spin coated on 
the $Cr_{2}O_{3}$ layer. This procedure is used to make a uniform 
thin layer of photoresist on the substrate by rotating the substrate 
at high speed in order to spread the resist by centrifugal force.

\item Exposure: the micro-structure pattern is transferred to the 
photoresist using a laser(412nm wavelength).  
Development: the photoresist (positive type) is removed.

\item The substrate is wet etched in an acid etch solution 
in a thin polyvinyl container.

\item The resist is stripped using a remover.

\end{enumerate}

\subsection{Masks on substrate}

Next, the masks on substrates were manufactured.
$\phi$ 30mm BK7 glass substrates, and 50mm-square Si 
and Ge substrates were used.
The detailed manufacturing process is shown below, and also in Fig.\ref{fig4}.
Manufacture of the two masks, \#SA004S and \#SA008S, failed and 
therefore these masks were not provided to the microscope check 
in the laboratory tests described in the nest section. 
The manufactured masks are shown in Fig.\ref{fig6}.

\begin{enumerate}

\item 
$\phi$ 30mm BK7 glass, 50mm-square Si, and 50mm-square Ge
are used as the substrates

\item Aluminum with thicknesses of 0.1/0.2/0.4/0.8/1.6$\mu$m is 
      deposited on the substrates by EB vapor deposition, a process 
      in which the aluminum is heated by an electron beam.

\item Photoresist with a thickness of 0.1 $\mu$m is spin coated on the Al.

\item Exposure: the pattern is transferred from the photo-mask 
      to the photo-resist by UV light (365nm wavelength) exposure. 
      Development: The photoresist (positive type) is removed.

\item The substrate is wet etched in an acid etch solution 
      in a thin polyvinyl container.

\item The photoresist is stripped by dipping in acetone.

\end{enumerate}

\subsection{Free-standing masks}

Lastly, the free-standing masks were manufactured.
Details of the manufacturing process of free-standing 
masks of Cu are given below, and also in Fig.\ref{fig5}.
The manufacturing process of free-standing masks of Ni are similar.
The manufactured masks are shown in Fig.\ref{fig6}.

\begin{enumerate}

\item A special 4-inch (101.6mm) square glass plate is used 
as the temporary substrate

\item A sacrificial release layer with a thickness of 1 $\mu$m is spin-coated on the surface.

\item A seed layer of Cu (0.5 $\mu$m thickness) is deposited by EB vapor deposition on the release layer.

\item Resist (with a goal of 10 $\mu$m thickness) is spin-coated for plating on the Cu substrate.

\item Exposure: the patterns are transferred from the photo-mask 
   to the photo-resist by exposure to UV light (365nm wavelength). 
   Development: the photo-resist is removed (only the illuminated 
   part is dissolved, i.e., the resist is positive type). 

\item Cu with goal thicknesses of 2/5/10/20 $\mu$m is grown 
   by electrolytic plating, a process in which metal ions in solution 
   are moved by an electric field to coat an electrode (i.e., the seed 
   layer on the substrate). 

\item The substrate is laminated with a dry film resist(100$\mu$m thickness) 
   to enable plating to be done for the support structure 
   around the border of the central micro-structure.

\item Exposure: the patterns are transferred from the 
   photomask to the photoresist by UV light exposure (365nm wavelength). 
   Development: the resist (negative type) is removed. 
   Only the illuminated parts of the dry film resist remain.

\item A thick layer of Cu with a goal thickness of 100$\mu$m 
   is deposited by electrolytic plating. 

\item The dry film resist is removed by dipping in acetone.

\item The photoresist is removed by dipping in acetone.

\item The Cu seed layer (0.5$\mu$m thickness) is etched by wet etching 
   the substrate in an acid etch solution in a thin polyvinyl container. 
   Not only the seed layer but also 0.5$\mu$m of Cu is removed in the process.   

\item The sacrificial release layer is removed by soaking in acetone.

\item The substrate is rinsed in isopropyl alcohol, then dried naturally.

\end{enumerate}

Achieved thickness of the holder area of the free-standing masks 
is $\sim$100$\mu$m and $\sim$170$\mu$m for Cu and Ni masks, respectively.

This work, comparative study of the mask manufacture, has been quested 
for several years(e.g., \citet{Enya2008cospar}; \citet{Enya2011b}).
A free standing mask of an early generation was used in \citet{Kotani2010}, 
and improvement of the manufacture process has been continued. 
Recently, a free-stainding mask having same specification with \#FC020
was adopted in \citet{Haze2012}. 
It should be noted that there are confidential details 
in the manufacturing processes, which are not described 
in this section explicitly.

\section{Laboratory tests}

\subsection{Microscope check}

The manufactured masks were checked with a digital optical 
microscope, VHX-90 made by KEYENCE Co.
Fig.\ref{fig7} and Fig.\ref{fig8} 
show examples of the microscope images, 
cases for a mask on a BK7 glass substrate, \#SA016B, 
and for a free-standing mask, \#FC020, 
respectively.
Since the role of the microscope check in this work is a qualifying 
round for the coronagraphic experiments, simply the topology 
of the mask was confirmed, rather than quantifying the imperfectness 
of the shape of many masks, i.e., we checked following:
1) All the holes in the design were reproduced in the manufactured masks. 
2) There were no holes unexpected in the design, in the manufactured mask.
3) All the holes in the design were separate from each other 
in the manufactured mask.
All the masks, except \#SA004S and \#SA008S for which the 
manufacturing process had failed, passed the microscope check,

\subsection{Cooling tests}

The masks on Si and Ge substrate are designed to be operated at 
cryogenic temperature, and cooling them to low temperature could 
potentially delaminate them due to mechanical stresses induced 
by mismatch of the coefficient of thermal expansion
between layers. 
We therefore carried out cooling tests for the masks on Si and Ge substrate.
All masks were installed onto the cold worksurface  of a cryostat in vacuum.
The worksurface was connected to a liquid nitrogen tank with a thermal
strap, and cooled by thermal conduction.
The masks were cooled to $\sim$80K in 10hours
(to $\sim$100K in 2hours), and then warmed up to ambient 
temperature in 20hours. 
Lastly microscopic check was applied,
and it is confirmed that no delamination 
was found for all cooled masks.


\subsection{Coronagraphic experiments}

\begin{figure}[t]
  \begin{center} 
  \FigureFile(88mm,){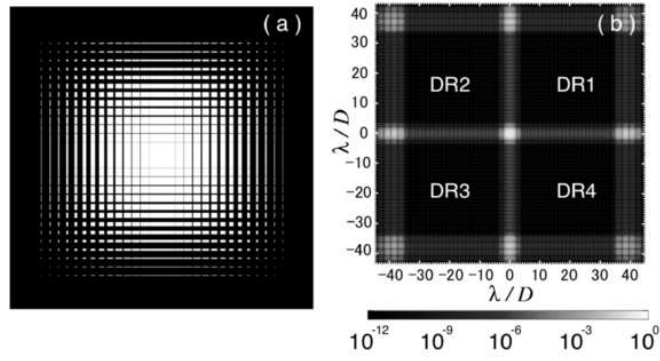}
 \end{center}
  \caption{Mask design. Left: design of the pupil mask.
Optical transmittances of the black and white areas are 0 and 1, respectively.
Right: simulated PSF. Four dark regions(DRs), DR1-DR4, 
are produced around the core of the PSF.
}\label{fig1}
\end{figure}

\begin{figure}[t]
  \begin{center} 
  \FigureFile(80mm,){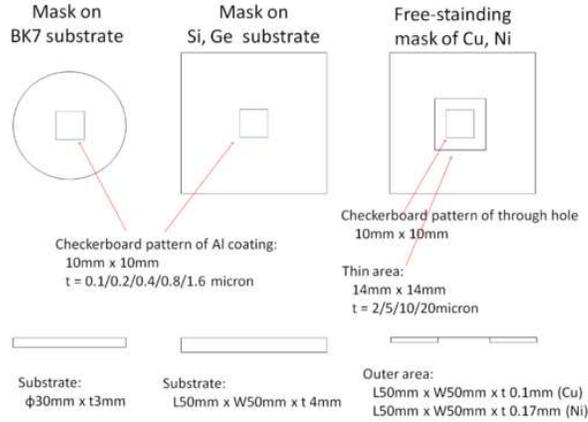}
  \end{center}
  \caption{
Geometry of the masks.
}\label{fig2}
\end{figure}

Coronagraphic experiments were carried out on all the masks 
with BK7 glass substrates and the free-standing masks.
The configuration of the experiment is shown in Fig.\ref{fig9}, 
which, except for the masks and their holders, is basically 
the same as the setup shown in section 2.4 of \citet{Haze2012}.
The optics were placed in a vacuum chamber, but vacuum pumping 
was not applied for the experiment presented in this paper.
Fig.\ref{fig7} and Fig.\ref{fig8} 
show examples of a mask installed in the holder, 
cases for a mask on a BK7 glass substrate, \#SA016B, 
for a free-standing mask, \#FC020, 
respectively.
All the coronagraphic images were taken using light passing 
through the mask (i.e., reflected light was not used).
A 632.8nm wavelength He-Ne laser was used as the light source, 
and this was introduced into the chamber through a single mode fiber.
All the focusing and collimation were executed with a plano-convex lens 
with anti-reflection coatings on both surfaces.
A CCD camera set in the chamber was used to take coronagraphic images.  
$\times$3.4 relay optics were set after the focal plane mask 
to obtain proper image sizes. 
To realize high dynamic range measurements, the cores 
and the DRs of the coronagraphic PSFs were taken separately. 
For each masks, we evaluated only the DR3 of the four DRs 
shown in Fig.\ref{fig1}
because of consideration for the efficiency of the experiment in this work.
The core images were taken with exposure times of 0.03/0.3/3 seconds 
using two ND filters with a total optical density of 4. 
The DR of the coronagraphic image was observed with a 300s exposure 
using a square hole focal-plane mask, without the ND filters. 
For all images, dark frames were taken with the same configuration, 
with the same exposure time, but with the light source turned off.
The dark frame was subtracted from the corresponding coronagraphic image, 
and then the raw image of the coronagraphic PSF was obtained.

The observed coronagraphic PSFs for a mask on a BK7 glass substrate, \#SA002B, 
are shown in Fig.\ref{fig10}, in which the left and the right panels are 
the image including the core, and the high sensitivity
image of the DR, respectively.
The observed coronagraphic PSFs taken with a free-standing mask of Cu, \#FC100, 
are also shown in Fig.\ref{fig11}.
In both cases, the observed core images are quite similar to 
the ones expected from the design presented in Fig.\ref{fig1}.
On the other hand, the observed dark images are filled 
with irregular speckle patterns.
This feature, core images expected from the design and the DR 
filled with speckle, was commonly found in 
all the coronagraphic PSFs obtained in this work.
Diagonal profiles of the coronagraphic images obtained from the masks 
with BK7 glass substrates and the free-standing Cu and Ni masks 
are presented in Fig.\ref{fig12}.

For all the masks tested in the coronagraphic experiments, 
the contrast was derived as the intensity ratio between the peak 
of the core and the linear average of the DR.
The contrasts obtained are presented in Table.\ref{table2}.

\section{Discussion and Summary}

The contrasts obtained are distributed from 5.3$\times10^{-8}$ to 2.1$\times10^{-7}$.
It should be noted that all the contrasts exceed the 
goal set at the beginning of this work, $10^{-6}$. 
The average contrasts were 8.4$\times10^{-8}$, 1.2$\times10^{-7}$, 
and 1.2$\times10^{-7}$ for the masks on BK7 glass substrates, 
and the free-standing masks of Cu and Ni, respectively.
The average contrast for the masks on BK7 glass substrates 
is higher than those of the free-standing masks.
However, because of the dispersion of the data, the statistical 
significance of this is not valid.
Also, significant correlation is not concluded between the contrast 
and the mask thickness for each of the three types of mask.

Because the contrast in the design is 10$^{-10}$, it is obvious 
that there is a practical limiting factor that gives rise 
to the speckle patterns observed in the DRs. 
However, identification of this limiting factor was not easy.
Using a mask having same specification with \#FC020, 
detail study about the limiting factor
was performed as shown in  \citet{Haze2012}: 
For example, mask rotation methods were tested.
In these tests, 
it is expected that the speckle patterns are rotated 
with the mask if the speckle patterns were produced 
simply by error of the mask shape.
However, less correlation was confirmed between the
speckle pattern before and after the mask rotation. 
Finally suggested candidates of the limiting factor
are imperfectness of incident
beam(e.g., wavefront error, inhomogeneity of amplitude,
and so on) and error in repeatability of the mask 
position before and after the rotation.
Influence of instability of the experimental system
is also suggested.
For more detail, please see \citet{Haze2012}.

The wavefront error, can be corrected by deformable mirrors
\citet{Trauger2007} pioneered ultra high contrast 
using the wavefront control with High Contrast Imaging Testbed.
Using one of early generation of our free-standing mask, 
\citet{Kotani2010} demonstrated improvement of the contrast
by factor of $\sim$100 at a part of the dark region
close to $IWA$ in the air with a visible laser.
For the use in infrared wavelength region,
development of cryogenic deformable mirror is ongoing.
Actuation of a proto-type of Micro Electro Mechanical Systems(MEMS) 
deformable mirror with 32 actuators were demonstrated 
at $\sim$95K(\cite{Enya2009}).
Toughness tests, vibration tests and rapid pumping tests 
were also carried out for the proto-type(\cite{Enya2011a}).

Coronagraphic experiments for the masks on Ge and Si substrates 
were not carried out in this work since visible light was used as 
the light source and the experiments were carried out at ambient temperature in air.
Important future work is to demonstrate the coronagraphic performance 
directly in the mid-infrared wavelength
region at cryogenic temperatures in vacuum.
Following results, indirectly,  suggest applicability
of the masks on Ge and Si substrates:
1) The high contrast was achieved using the masks on BK7 glass 
substrates in this work.
2) The masks on Ge and Si substrates and
the masks on BK7 glass substrates 
were manufactured using same process.
3) The masks on Ge and Si substrates survived the cooling tests.
It is also important to evaluate coronagraph
using the free-standing masks in infrared at cryogenic temperature.
Finally, performance of the all the masks for an infrared
coronagraph should be compared.

Only one mask design, a checkerboard type without pupil obscuration, 
was used in this work to compare the various manufacturing processes.
On the other hand, recent progress in mask design allows the binary 
pupil mask coronagraph to be applied to a normal telescopes with pupil 
obscuration, which is not specially designed for a coronagraph.
\citet{Carlotti2011} presented a 2-dimensionally optimized pupil 
design which provides the ultimate efficiency possible in terms of 
throughput for a pupil coronagraph mask. 
An integral 1-dimensional coronagraph pupil mask also gives a higher 
throughput than conventional ones, and a generalized design of the 
dark region at the focal plane was introduced to realize a more 
efficient distribution of the $IWA$, $OWA$, and contrast at the 
focal plane (\cite{EnyaAbe2010}; \cite{Enya2011a}).
As a result, these high contrast masks have the potential to cover 
the needs of coronagraphs for ground-based telescopes(e.g., current 8-10m 
class telescopes like SUBARU, and larger future ones such as TMT, EELT), 
and space telescopes (e.g., JWST, SPICA) over a wide wavelength region. 
Indeed, the use of a binary pupil mask coronagraph is planned for 
the SPICA Coronagraph Instrument(SCI), for which the results of this 
work are quite encouraging.
Because of less wavelength dependence of binary pupil mask coronagraphs, 
it would be worthy
to evaluate benefit of applying binary pupil masks for  
instruments for SPICA for longer wavelength;
Mid-infrared Camera and Spectrometer(MCS; \cite{Kataza2010}) and/or 
SPICA FAR-infrared Instrument(SAFARI; e.g., \cite{Goicoechea2012}).

In the work presented in this paper, we carried out a comparative 
study of the manufacturing processes of binary pupil masks for coronagraphs. 
Both masks on substrates and free-standing masks were manufactured 
with various materials and thicknesses.
Coronagraphic experiments in the visible light region confirmed 
the high contrast, in which obtained average contrasts 
were 8.4$\times10^{-8}$, 1.2$\times10^{-7}$, and 1.2$\times10^{-7}$ 
for the masks on BK7 substrates, the free-standing copper masks, 
and the free-standing nickel masks, respectively.
Significant correlation was not concluded between the contrast 
and the mask properties.
We consider such masks have the potential to cover needs of 
coronagraphs for various telescopes.

\bigskip

We are grateful to the all pioneers in this field, 
especially to R. J. Vanderbei.
The work is supported by the Japan Society for the Promotion of Science, 
the Ministry of Education, Culture, Sports, Science and Technology of Japan, 
and the Japan  Aerospace Exploration Agency. 
We thank A. Suenaga, T. Ishii and their colleagues 
in Houwa-sangyo Co. and Photo-precision. Co.
We also thak to referee's fruitful comments for this paper.
Lastly, we would like to express special thanks to S. Tanaka, 
and wish him well in his current field.



\newpage

\begin{table*}[ht]
  \begin{center}
  \caption{Summary of the mask specifications.}
  \label{table1}
     \begin{tabular}{llllll}
      \hline 
      \hline 
      Type   & No. &   Mask     &    Thickness($\mu$m)   & Substrate  & Note  \\
      \hline 
      on substrate & \#SA001B    & Al  &    0.1  &  BK7   & AR$^*$     \\
                     & \#SA002B    & Al  &    0.2  &  BK7   & AR$^*$        \\
                     & \#SA004B    & Al  &    0.4  &  BK7   & AR$^*$        \\
                     & \#SA008B    & Al  &    0.8  &  BK7   & AR$^*$        \\
                     & \#SA016B    & Al  &    1.6  &  BK7   & AR$^*$        \\
                     & \#SA001S    & Al  &    0.1  &  Si    & CL$^\dag$          \\
                     & \#SA002S    & Al  &    0.2  &  Si    & CL$^\dag$          \\
                     & \#SA004S    & Al  &    0.4  &  Si    & Failed$^\ddag$    \\
                     & \#SA008S    & Al  &    0.8  &  Si    & Failed$^\ddag$    \\
                     & \#SA016S    & Al  &    1.6  &  Si    & CL$^\dag$          \\
                     & \#SA001G    & Al  &    0.1  &  Ge    & CL$^\dag$     \\
                     & \#SA002G    & Al  &    0.2  &  Ge    & CL$^\dag$     \\
                     & \#SA004G    & Al  &    0.4  &  Ge    & CL$^\dag$     \\
                     & \#SA008G    & Al  &    0.8  &  Ge    & CL$^\dag$     \\
                     & \#SA016G    & Al  &    1.6  &  Ge    & CL$^\dag$     \\
      \hline 
      Free-standing  & \#FC020     & Cu  &    2 &  ---    &      \\
                     & \#FC050     & Cu  &    5 &  ---    &    \\
                     & \#FC100     & Cu  &   10 &  ---    &    \\
                     & \#FC200     & Cu  &   20 &  ---    &    \\
                     & \#FN020     & Ni  &    2 &  ---    &    \\
                     & \#FN050     & Ni  &    5 &  ---    &    \\
                     & \#FN100     & Ni  &   10 &  ---    &    \\
                     & \#FN200     & Ni  &   20 &  ---    &    \\
      \hline 
      \end{tabular}
  \end{center}
    $^*$ Anti-reflection coatings were applied to both sides.\\
    $^\dag$ Cooling tests were applied.\\
    $^\ddag$ Manufacturing failed.\\
\end{table*}

\begin{table*}[t]
  \begin{center}
  \caption{Contrast obtained by experiment.}
  \label{table2}
     \begin{tabular}{llllll}
      \hline 
      \hline 
      No.  & Contrast  &  No.   & Contrast  &  No.   & Contrast\\  
      \hline 
    \#SA001B  & 6.6$\times10^{-8}$  &  \#FC020   & 2.1$\times10^{-7}$  &  \#FN020   &  1.1$\times10^{-7}$  \\
    \#SA002B  & 1.1$\times10^{-7}$  &  \#FC050   & 5.3$\times10^{-8}$  &  \#FN050   &  8.5$\times10^{-8}$  \\
    \#SA004B  & 5.9$\times10^{-8}$  &  \#FC100   & 1.1$\times10^{-7}$  &  \#FN100  &  1.2$\times10^{-7}$  \\
    \#SA008B  & 7.3$\times10^{-8}$  &  \#FC200   & 9.3$\times10^{-8}$  &  \#FN200  &  1.8$\times10^{-7}$  \\
    \#SA016B  & 1.1$\times10^{-7}$  &          &                    &          &   \\
      \hline 
     Average &     8.4$\times10^{-8}$      &     &  1.2$\times10^{-2}$  &  &            1.2$\times10^{-7}$     \\
      \hline 
      \end{tabular}
  \end{center}
\end{table*}


\clearpage

\clearpage
  \begin{center}
\begin{figure}[ht]
  \begin{center}
     \FigureFile(80mm, 240mm){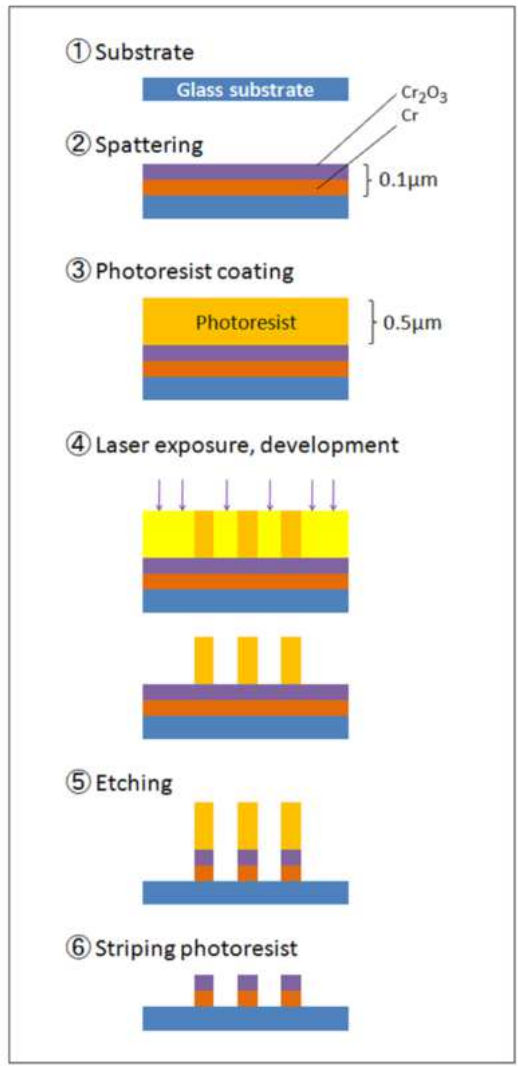}
   \end{center}
  \caption{Manufacturing process of the photomask.
}\label{fig3}
\end{figure}
\end{center}

\begin{figure}[h]
  \begin{center}
     \FigureFile(80mm, 240mm){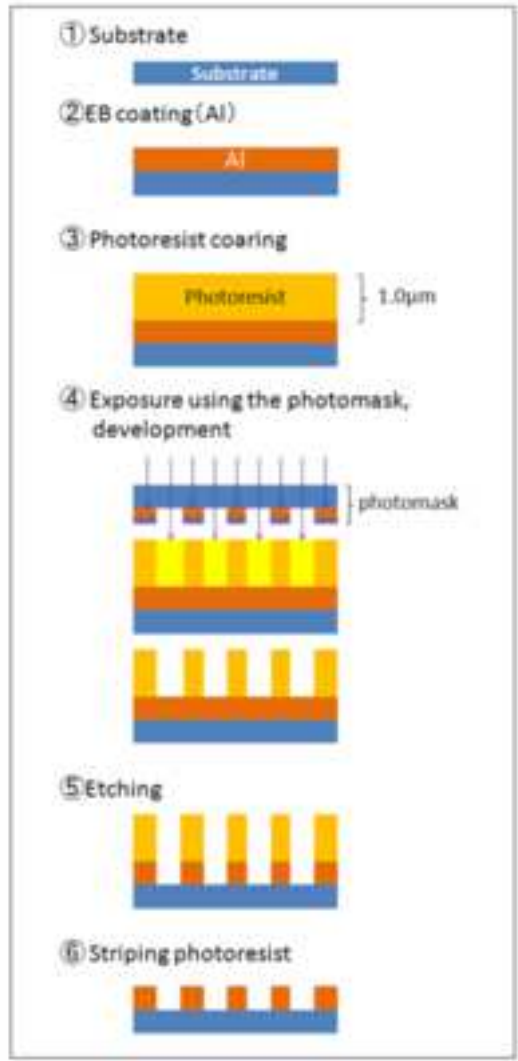}
  \end{center}
  \caption{
Manufacturing process of masks on substrates.
}\label{fig4}
\end{figure}

\clearpage
\begin{figure*}[ht!]
  \begin{center}
     \FigureFile(150mm, ){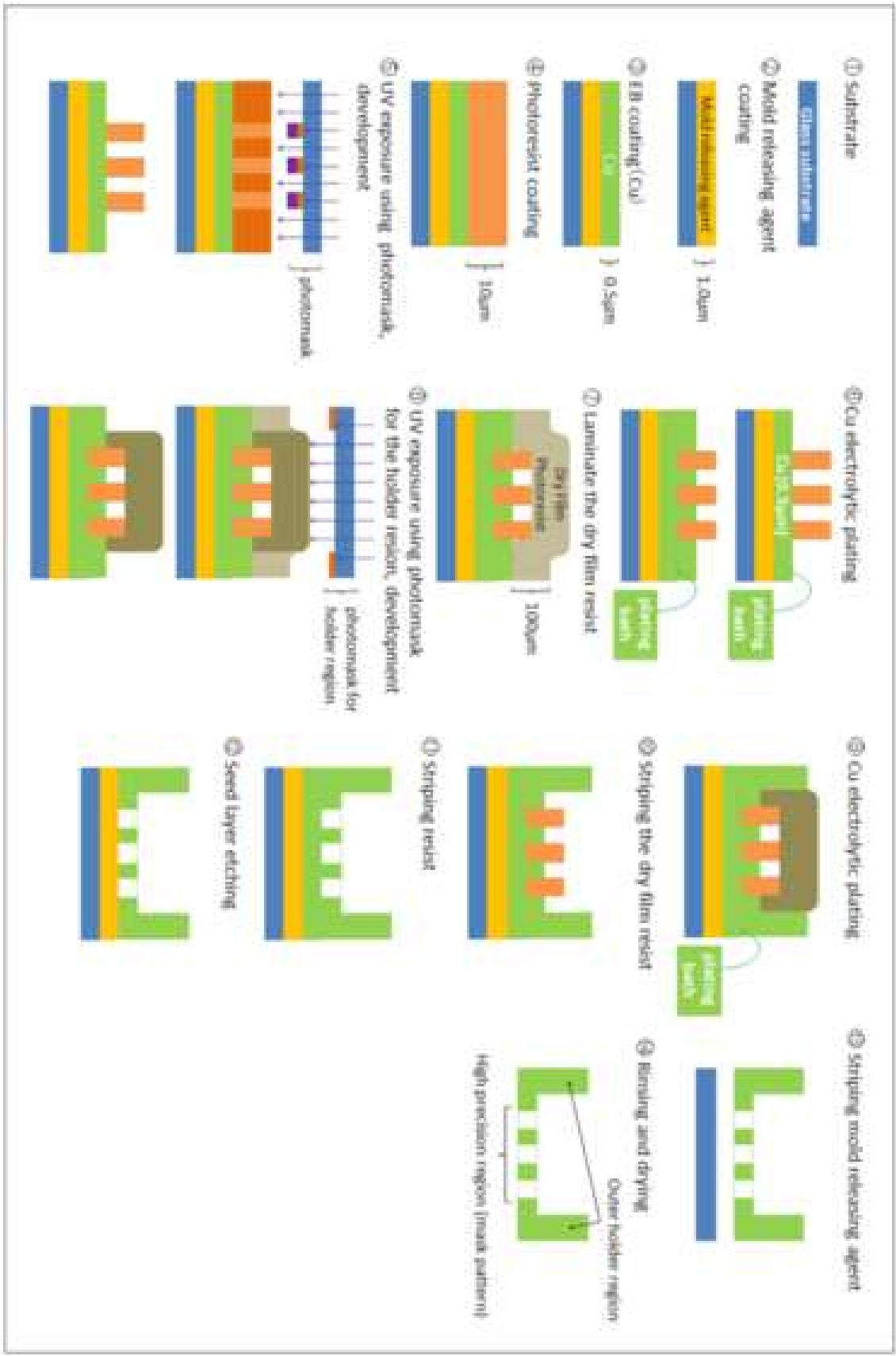}
  \end{center}
  \caption{
Manufacturing process of free-standing masks of Cu. 
Manufacturing process of free-standing masks of Ni are similar. 
}\label{fig5}
\end{figure*}

\begin{figure*}[h!]
  \begin{center}
     \FigureFile(140mm,){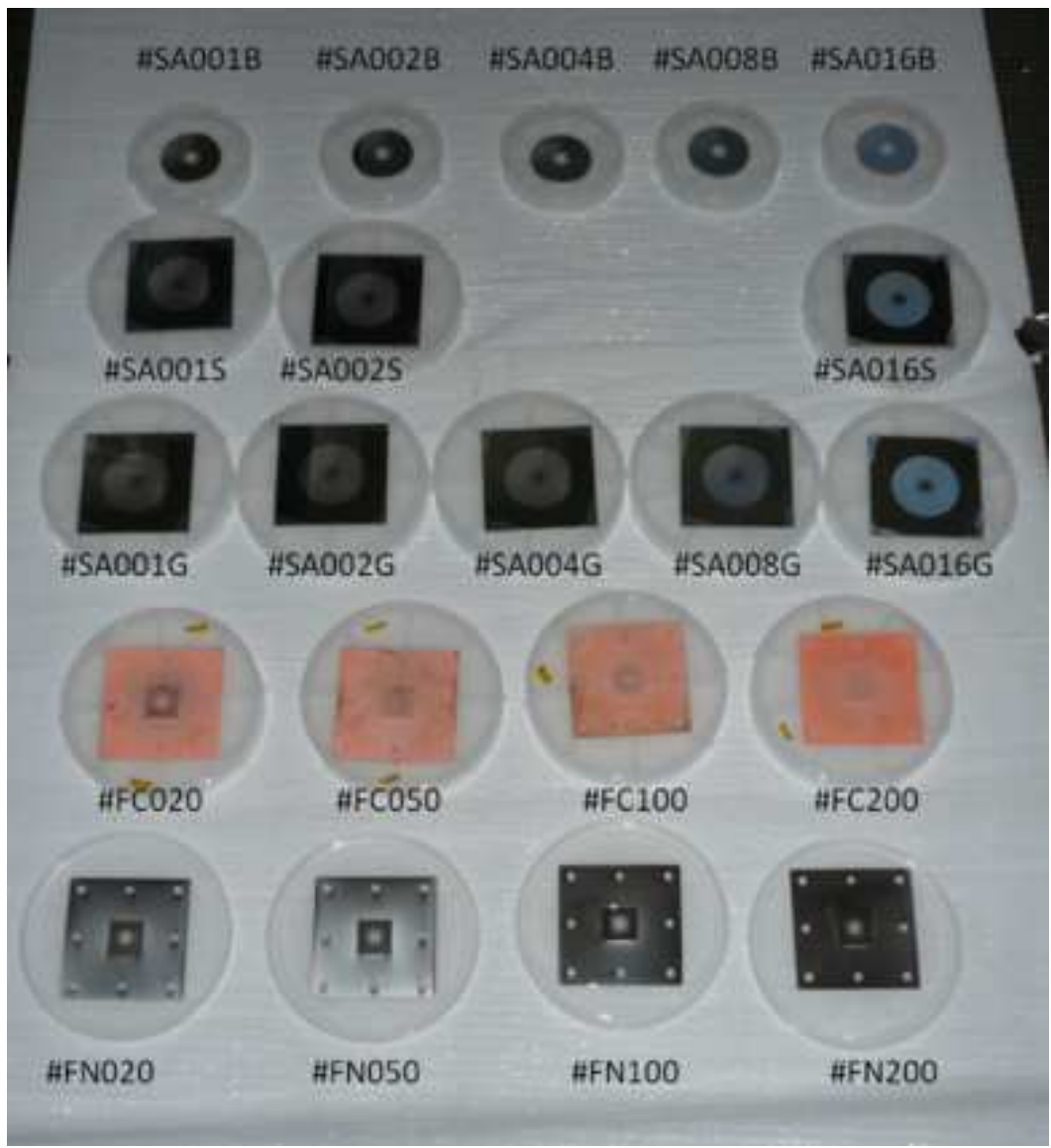}
  \end{center}
  \caption{
Manufactured masks.
}\label{fig6}
\end{figure*}

\clearpage
\begin{figure*}[ht]
  \begin{center}
     \FigureFile(150mm, ){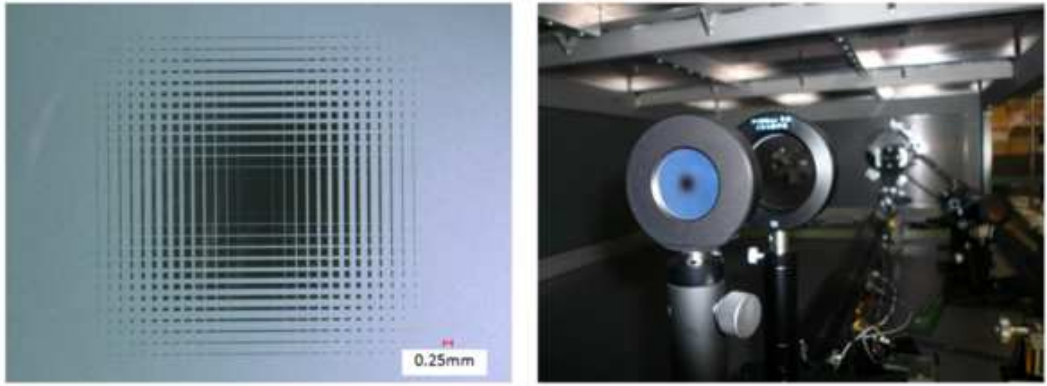}
  \end{center}
  \caption{
Left: microscope image of a mask on a BK7 glass substrate, \#SA016B.
Right: the mask installed in the holder.
}\label{fig7}
\end{figure*}

\begin{figure*}[ht]
  \begin{center}
     \FigureFile(150mm, ){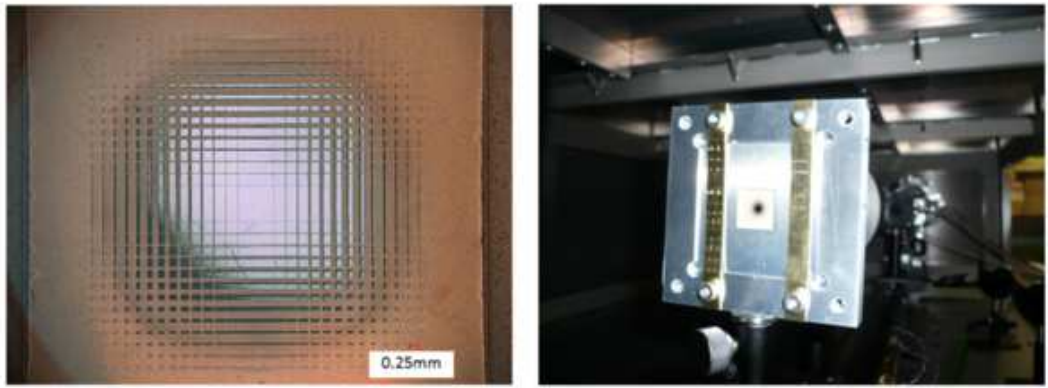}
  \end{center}
  \caption{
Left: microscope image of a free-standing mask of Cu, \#FC020.
Right: mask installed in the holder.
}\label{fig8}
\end{figure*}

\clearpage
\begin{figure}
  \begin{center}
     \FigureFile(88mm, ){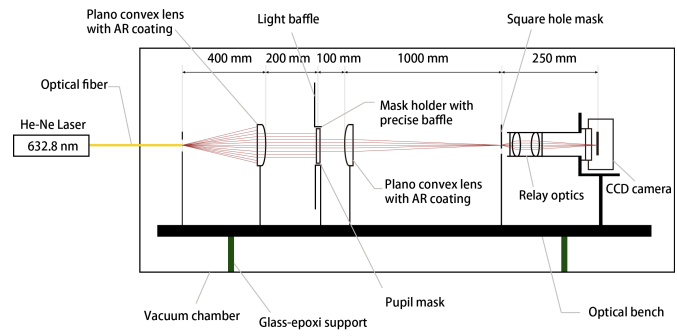}
  \end{center}
  \caption{
Configuration of the coronagraphic experiment.
}\label{fig9}
\end{figure}

\clearpage
\begin{figure*}
  \begin{center}
     \FigureFile(170mm, ){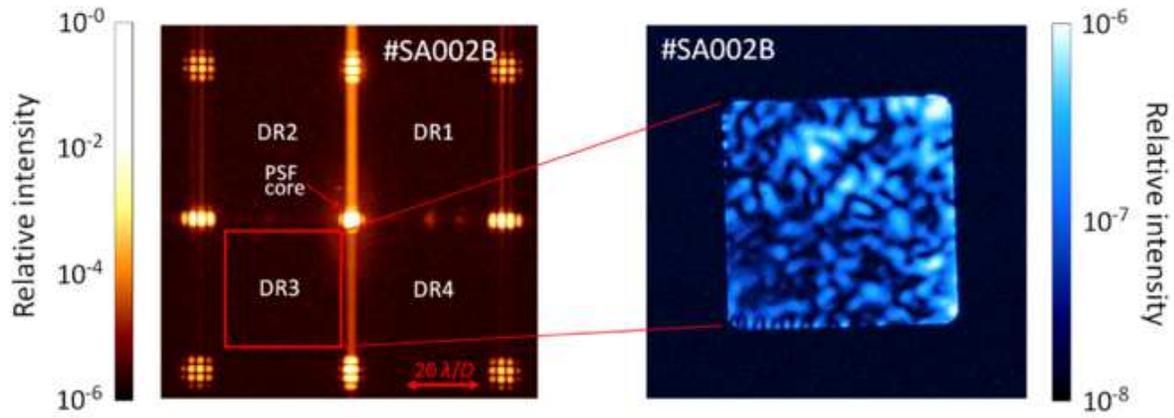}
  \end{center}
  \caption{
Coronagraphic image taken with a mask on a BK7 glass substrate, \#SA002B.
Left: image includes the core of the PSF.
Right: high sensitivity image of the DR taken with a square hole mask.  
}\label{fig10}
\end{figure*}

\begin{figure*}
  \begin{center}
     \FigureFile(170mm, ){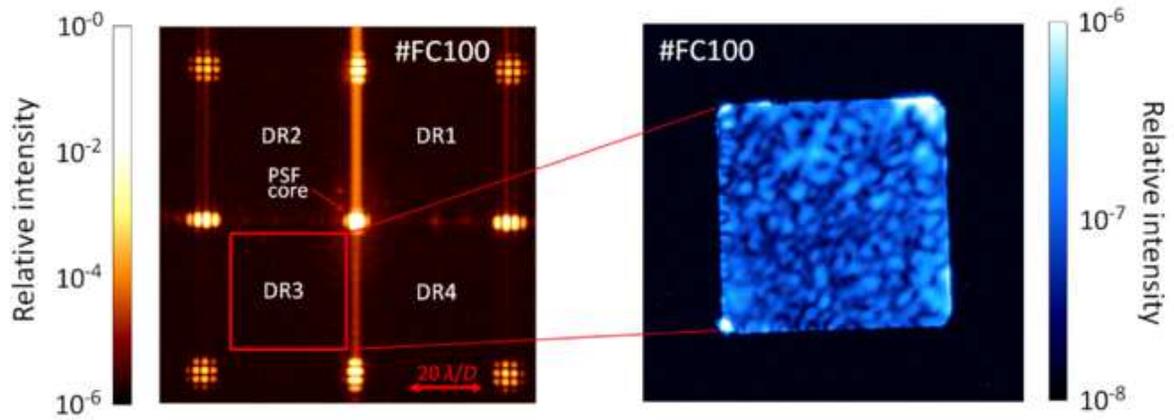}
  \end{center}
  \caption{
Coronagraphic image taken with a free-standing mask of Cu, \#FC100.
Left: image includes the core of the PSF.
Right: high sensitivity image of the DR taken with a square hole mask. 
}\label{fig11}
\end{figure*}

\clearpage
\begin{figure*}[ht!]
  \begin{center}
     \FigureFile(88mm, ){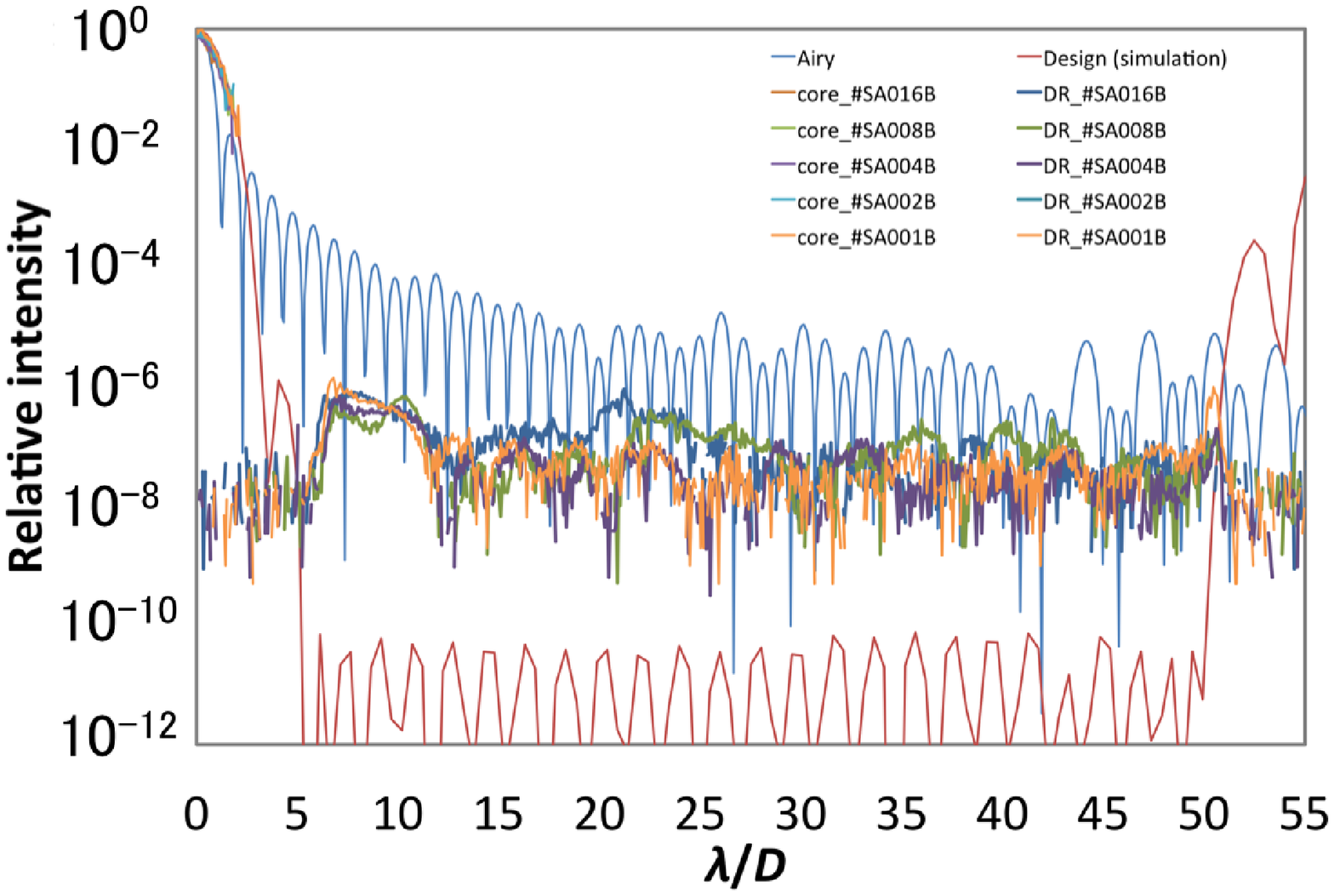}
     \FigureFile(88mm, ){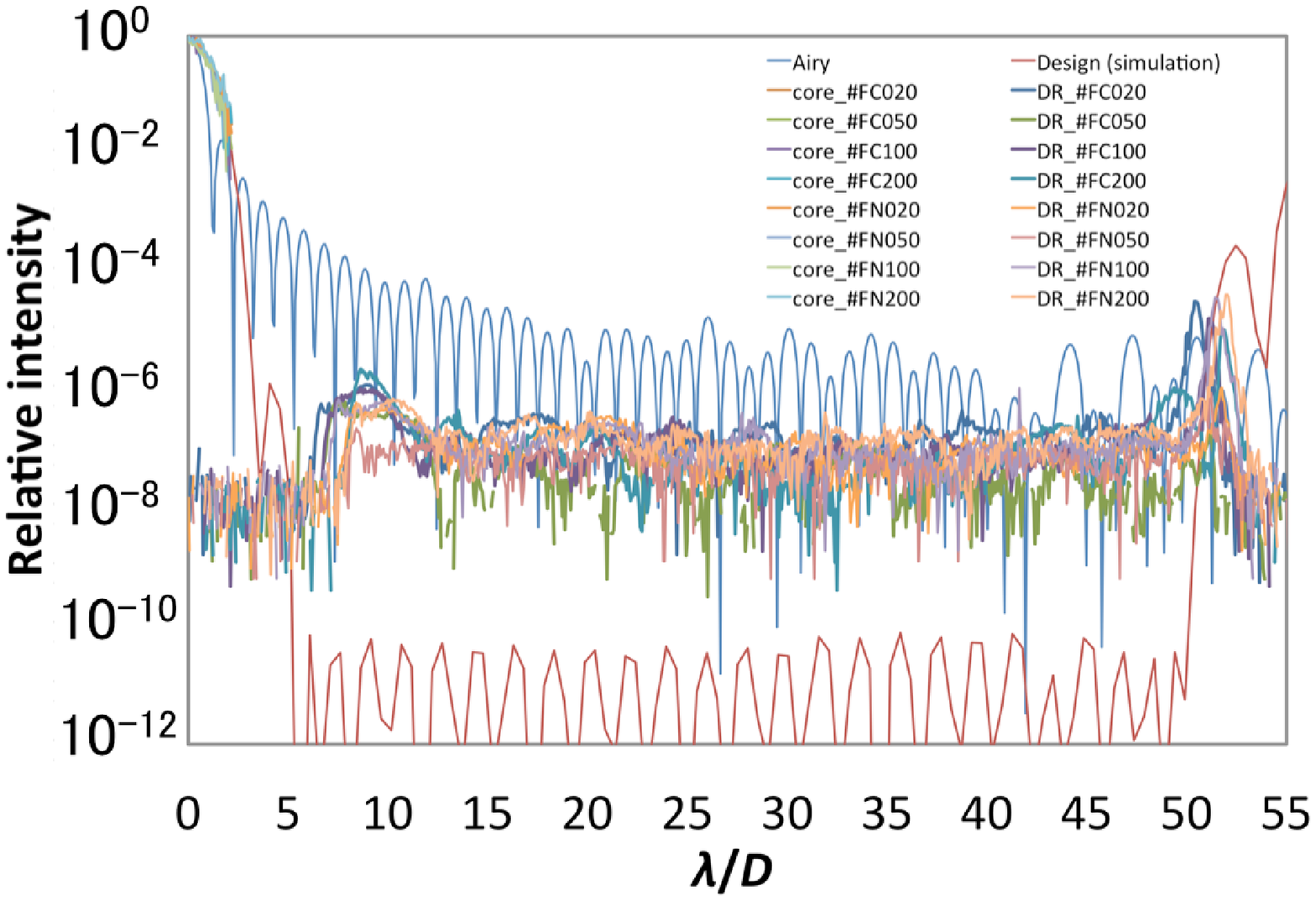}
  \end{center}
  \caption{
Left: diagonal profiles of coronagraphic images obtained with masks on 
BK7 substrate.
Right: diagonal profiles of coronagraphic images obtained 
 with free-standing masks.
}\label{fig12}
\end{figure*}


\begin{thebibliography}{}



\bibitem[Carlotti, Vanderbei, \& Kasdin(2011)]{Carlotti2011}
Carlotti, A., Vanderbei, R., \& Kasdin, N. J.,\
2011,  arXiv:1108.4050


\bibitem[Charbonneau et al.(2000)]{Charbonneau2000}
      Charbonneau, D.,  Brown, T. M., Latham, D. W., \& Mayor, M.\ 
      2000, \apjl, 529, 45


	
\bibitem[Belikov et al.(2007)]{Belikov2007}
   Belikov, R., Give'on, A., Kern, B., Cady, E., Carr, M., 
   Shaklan, S., Balasubramanian, K., White, V., 
   Echternach, P., Dickie, M., Trauger, J., 
   Kuhnert, A., Kasdin, N. J.
  2007, proc. of SPIE, 6693,  66930Y 

	

\bibitem[Enya et al.(2011a)]{Enya2011a}
Enya, K.,  Abe, L.,  Takeuchi, S., Kotani, T., \& Yamamuro, T.,\
2011, Proc. of SPIE, 8146, 81460Q

\bibitem[Enya et al.(2011b)]{Enya2011b}
   Enya, K. et al.\
   2011, Advances in  Space Research,  48, 323



\bibitem[Enya et al.(2010)]{Enya2010}
   Enya, K., \& SPICA working~group\
   2010,  Advances in  Space Research,  45, 979


\bibitem[Enya \& Abe(2010)]{EnyaAbe2010}
Enya, K. \& Abe, L.\ 
2010, \pasj, 62, 1407



\bibitem[Enya et al.(2009)]{Enya2009}
Enya, K., Kataza, H., \& Bierden, P.\
2009, \pasp, 121, 260


\bibitem[Enya et al.(2008)]{Enya2008}
   Enya, K., Abe, L., Tanaka, S., Nakagawa, T., Haze, K., Sato, T., Wakayama, T.
   2008, \aap, 461, 783


\bibitem[Enya(2008)]{Enya2008cospar}
   Enya, K.\
   37th COSPAR Scientific Assembly. Held 13-20 July 2008, 
   in Montr\`el, Canada., p.812


\bibitem[Enya et al.(2007)]{Enya2007}
   Enya,~K.,  Tanaka,~S., Abe,~L., \& Nakagawa,~T.\
   2007, \aap, 480, 899



\bibitem[Goicoechea et al.(2012)]{Goicoechea2012}
   Goicoechea, J. R., Roelfsema, P. R., 
   Jellema, W., \& Swinyard, B. M.\
   proc. of the 280th Symposium of the IAU held in Toledo, 
   Spain, May 30-June 3, 2011, \#179


\bibitem[Haze(2012)]{Haze2012}
   Haze, K.\
   2012, PhD. thesis, arXiv1112.6301H

\bibitem[Haze(2011)]{Haze2011}
   Haze, K., Enya, K., Abe, L., Kotani, T., Nakagawa, T., 
   Sato, T., Yamamuro, T.\
   2011, \pasj, 63, 873

\bibitem[Haze(2009)]{Haze2009}
Haze, K., Enya, K., Abe, L., Tanaka, S., Nakagawa, T., 
Sato, T., Wakayama, T., Yamamuro, T.\
2009,  Advances in  Space Research, 43, 181

\bibitem[Jacquinot \& Roizen-Dossier(1964)]{Jacquinot1964}
Jacquinot, P., \& Roizen-Dossier, B. 1964, Prog. Opt., 3, 29


\bibitem[Kalas et al.(2008)]{Kalas2008}
Kalas, P., Graham, J. R., Chiang, E., Fitzgerald, M. P., \& Clampin,~M.\ 
2008, Science, 322, 1345



\bibitem[Kataza et al.(2010)]{Kataza2010}
Kataza, H., Wada, T., Ikeda, Y., Fujishiro, N., 
Kobayashi, N., \& Sakon, I.\
2010, proc. of SPIE, 7731, 77314A



\bibitem[Kasdin et al.(2005a)]{Kasdin2005a}
    Kasdin, N. J., Vanderbei, R. J., Littman, M. G. \& Spergel, D. N.\ 
    2005, \ao, 44, 1117


\bibitem[Kasdin et al.(2005b)]{Kasdin2005b}
Kasdin, N. J., Belikov, R., Beall, J., Vanderbei, R. J.\ 
Littman, M. G., Carr, M., \& Give'on, A.\  
2005, Proc. of SPIE, 5905, 128


\bibitem[Lagrange et al.(2010)]{Lagrange2010}
Lagrange, A.-M., Bonnefoy, M., Chauvin, G., Apai, D., 
Ehrenreich, D., Boccaletti, A., Gratadour, D., Rouan, D., 
Mouillet, D., Lacour, S., \& Kasper, M.\
2010, Science, 329, 57L



\bibitem[Lyot(1939)]{Lyot1939}
Lyot, B.\
1939, \mnras, 99, 580



\bibitem[Kotani(2010)]{Kotani2010}
Kotani, T., Enya, K., Nakagawa, T., Abe, L., Miyata, T., 
Sako, S., Nakamura, T., Haze, K., Higuchi, S., Tange, Y.\
2010,
Proc. of a workshop held 14 to 18 September 2009 in Barcelona, 
Spain. Edited by Vincent Coud\'e  du Foresto, Dawn M. Gelino, 
and Ignasi Ribas. San Francisco: Astronomical Society of the Pacific, p.477

\bibitem[Marois et al.(2008)]{Marois2008}
Marois, C., Macintosh, B., Barman, T., 
Zuckerman, B.,  Song, I., Patience, J.,  Lafreni\`ere, D., \& Doyon,~R.\
2008, Science, 322, 1348



\bibitem[Mayor \& Queloz(1995)]{Mayor1995}
     Mayor,~M., \& Queloz,~D.\
     1995, \nat, 378, 355


\bibitem[Nakagawa et al.(2009)]{Nakagawa2009}
   Nakagawa,~T., \& Spica Team\
   2009, SPICA joint European/Japanese Workshop, held 6-8 July, 
   2009 at Oxford, United Kingdom. Edited by A.~M.~Heras, 
   B. M. Swinyard, K. G. Isaak, and J. R. Goicoechea. 
   EDP Sciences, 2009, p.01001


\bibitem[Spergel(2001)]{Spergel2001}
Spergel, D. N., 2001, arXiv:astro-ph/0101142



\bibitem[Tanaka et al.(2006)]{Tanaka2006}
    Tanaka,~S., Enya,~K., Nakagawa,~T., Kataza,~H., \& Abe,~L.\
    2006, \pasj, 58, 627



\bibitem[Traub \& Jucks (2002)]{Traub2002}
Traub,~W.~A., \& Jucks,~K.~W.\
2002, Astro-ph/0205369


\bibitem[Trauger \& Traub(2007)]{Trauger2007}
    Trauger,~J.~T., \& Traub,~W.~A.\
    2007, \nat, 446, 771


\bibitem[Vanderbei, Kasdin, \& Spergel(2004)]{Vanderbei2004}
    Vanderbei,~R.~J., Kasdin,~N.~J. \& Spergel,~D.~N.\
    2004, \apj, 615, 555

\bibitem[Vanderbei, Kasdin, \& Spergel(2003a)]{Vanderbei2003a}
    Vanderbei,~R.~J., Kasdin,~N.~J. \& Spergel,~D.~N.\
    2003, \apj, 590, 593


\bibitem[Vanderbei, Spergel, \& Kasdin(2003b)]{Vanderbei2003b}
Vanderbei, R. J., Spergel, D. N., and Kasdin, N. J., 
2003, \apj, 599, 686


\bibitem[Vanderbei(1999)]{Vanderbei1999}
   Vanderbei,~R.~J.\
   1999, Optimization methods and software, 11, 451








\end{thebibliography}
\end{document}